\documentstyle [12pt,a4]{article}

\newcommand{\be}{\begin{equation}}
\newcommand{\ee}{\end{equation}}
\newcommand{\bea}{\begin{eqnarray}}
\newcommand{\ena}{\end{eqnarray}}
\newcommand{\vs}[1]{\rule[- #1 mm]{0mm}{#1 mm}}

\newcommand{\sm}[2]{\frac{\mbox{\footnotesize #1}\vs{-2}}
                   {\vs{-2}\mbox{\footnotesize #2}}}

\newcommand{\shalf}{\sm{1}{2}}

\newcommand{\eps}{\epsilon}

\newcommand{\dof}{\hat{s}l_q(2,{\bf C})}
\newcommand{\dom}{sl_q(2,{\bf C})}
\newcommand{\jn}{j \leq \frac{N-1}{2}}
\newcommand{\hr}{\hat{R}}
\newcommand{\vv}{v^{(1)}_c \otimes v^{(2)}_c}
\newcommand{\vvv}{v^{(2)}_c \otimes v^{(1)}_c}

\newcommand{\NP}[1]{Nucl.\ Phys.\ {\bf #1}}
\newcommand{\PL}[1]{Phys.\ Lett.\ {\bf #1}}

\newcommand{\CMP}[1]{Comm.\ Math.\ Phys.\ {\bf #1}}

\newcommand{\LMP}[1]{Lett.\ Math.\ Phys.\ {\bf #1}}

\begin{document}
\renewcommand{\thefootnote}{\fnsymbol{footnote}}
\newpage
\setcounter{page}{0}

\vs{15}

\begin{center}

{\large {\bf $R$-matrices for highest weight representations \\[.5cm]
of $\hat{s}l_q(2,{\bf C})$ at roots of unity}}\\[1.5cm]

{\large T.S. Hakobyan}\\
{\em Yerevan State University, Manukyan\\
st. 9, Yerevan, Armenia}\\[.5cm]

{\em and}\\[.5cm]

{\large A.G. Sedrakyan \footnote{permanent address : Yerevan Physics Institute,
Br. Alikhanian, st.2, Yerevan 36, Armenia}}\\
{\em{Laboratoire de Physique Th\'eorique}}
{\small E}N{\large S}{\Large L}{\large A}P{\small P}
\footnote{URA 14-36 du CNRS, associ\'ee \`a l'E.N.S. de Lyon et au L.A.P.P.
(IN2P3-CNRS) d'Annecy-le-Vieux}
\\
{\em Chemin de Bellevue BP 110, F - 74941 Annecy-le-Vieux Cedex,
France}\\[.5cm]

\end{center}
\vs{10}

\centerline{\bf Abstract}

\indent

We obtain the general formula for $R$-matrices of $\dof$ for the highest weight
representations both for general $q$ and for $q$ being a root of unity by
generalizing G. Gomez's and Sierra's one for semiperiodic representations of
$\dof$ at roots of unity. In a case of general $q$ we obtain direct matrix form
for the well known Jimbo's $R$-matrix. In a case of $q^N=1$ for semiperiodic
and spin $j < \frac{N-1}{2}$ representations we obtain the new $R$-matrices
which together with Jimbo's one obey the spectral parameter dependent
Yang-Baxter equations.

\vs{15}

\rightline{{\small E}N{\large S}{\Large L}{\large A}P{\small P}-A-406/92}
\rightline{October 1992}

\vfill

\newpage

\renewcommand{\thefootnote}{\arabic{footnote}}
\setcounter{footnote}{0}
\newpage

\indent

Quantum groups had been introduced by Drinfeld \cite{1} and Jimbo \cite{2}.
Shortly speaking, quantum group is some quasi-triangular Hopf algebra,
parametrized by some parameter $q$ \cite{1}. They play an important role in
two-dimensional integrable statistic models and conformal field theory.

For general $q$ irreducible representations of quantum algebras are in one to
one correspondence with the representations of the classical algebras
\cite{2,15}. For $q$ being a root of 1 this is not true. Irreducible
representations in this case had been considered in \cite{7,10,11}. They are
parametrized by some complex numbers.

Due to Hopf algebra structure one can consider an intertwiner on tensor
products of two irreducible representations \cite{1,2}. In case of general $q$
intertwiners satisfy Yang-Baxter equations \cite{1,3}, that is the
integrability
condition for the two dimensional systems.

For $q$ being a root of unity the standard expression of the universal
$R$-matrix \cite{1} is ill defined because of singularities. In \cite{9}
Bazhanov and Stroganov showed how the chiral Potts model \cite{17,18} is
related to the theory of cyclic representations of the quantum algebra $\dof$
at $q^N=1$. Intertwiners between semiperiodic representations of $\dom$ had
been considered in \cite{4}. In \cite{12,14} the intertwiners between
semiperiodic and spin $\jn$ representations, where $N$ is a minimal integer,
satisfying the condition $q^N=1$, had been considered. In \cite{4} Gomez and
Sierra found an interesting recursive formula for intertwiners of semiperiodic
representations, having a reflection symmetry.

In this article we generalize this formula for the case of affine $\dof$
algebra. It can be used for the highest weight representations of ordinary
$\dom$ algebra. In a particular case of general $q$ we obtain well known
Jimbo's $R$-matrix \cite{2}, written in the matrix form. In case of $q$ being
an odd root of 1 by acting our general formula to tensor product of
semiperiodic representation and spin $\jn$ representation, we obtain
baxterization of intertwiners, considered in \cite{12}. As a consequence we
obtain new solutions of spectral parameter dependent Yang-Baxter equations. We
show also that the intertwiners, considered in \cite{4,5}, are particular cases
of ones, obtained in \cite{6} by a different approach.

Let us recall the definition of the affine quantum algebra $\dof$. It is
generated by generators $e_i, f_i, h_i$, satisfying the following relations
\cite{2}:
\bea
[h_i,e_j] &=& a_{ij} \ e_j \ \ [h_i, f_j] = -a_{ij} \ f_j \ \ \ [h_i, h_j] =0
\nonumber \\
{[}e_i,f_j{]} &=& \delta_{ij} \ [h_j]_q:=\delta_{ij} \ \frac{q^{h_i} -
q^{-h_i}}{q-q^{-1}}
\label{eq:1}
\ena

(Here $a_{ij} = \left( \begin{array}{rr}
2 & -2 \\
-2 & 2
\end{array}
\right)$ is Cartan matrix of $\dof$, $q$ is a parameter)

\bea
&& \sum^{1-a_{ij}}_{\nu=0} (-1)^\nu \left[ \begin{array}{c}
1-a_{ij} \\
\nu
\end{array}
\right]_q
e_i^{1-a_{ij}- \nu} e_j e_i^\nu =0 \ \ \ (i \neq j) \nonumber \\
&& \sum^{1-1_{aj}}_{\nu=0} (-1)^\nu \left[ \begin{array}{c}
1-a_{ij} \\
\nu
\end{array}
\right]_q
f_i^{1-a_{ij}- \nu} f_j f_i^\nu =0 \ \ \ (i \neq j)
\label{eq:2}
\ena

\[
\mbox{(Here} \left[ \begin{array}{c}
m\\
n
\end{array}
\right]_q : = \frac{[m]!_q}{[m-n]_q ! [n]_q!}, \ \ \ \ [m]_q!:=[1]_q [2]_q ..
[m]_q).
\]

\indent

Following \cite{6} we take the extension of $\dof$ by central elements
$z^{\pm1}_i (i=0,1)$. There is the structure of Hopf algebra in this algebra:
\be
\begin{array}{ll}
\Delta (e_i) = e_i \otimes k^{-1}_i + z_i \cdot k_i \otimes e_i &
\Delta (k^{\pm 1}_i) = k^{\pm 1}_i \otimes k^{\pm 1}_i \nonumber \\
\Delta (f_i) = f_i \otimes k^{-1}_i + z^{-1}_i \cdot k_i \otimes f_i &
\Delta (z^{\pm 1}_i) = z^{\pm 1}_i \otimes z^{\pm 1}_i ,
\label{eq:3}
\end{array}
\ee
where $k_i:=q^{\frac{h_i}{2}}$.

Note, that in \cite{6} another comultiplication had been considered:
\be
\begin{array}{ll}
\tilde{\Delta} (e_i) = e_i \otimes k^{-\eps_i} + z_i \cdot q^{\eps_i}
\otimes e_i &
\tilde{\Delta} (q^{\eps_i}) = q^{\eps_i} \otimes q^{\eps_i} \nonumber \\
\tilde{\Delta} (f_i) = f_i \otimes q^{-\eps_{i+1}} + z^{-1}_i
\cdot q^{-\eps_{i+1}} \otimes f_i &
\tilde{\Delta} (z^{\pm 1}_i) = z^{\pm 1}_i \otimes z^{\pm 1} _i
\label{eq:4}
\end{array}
\ee
where $\eps_i (i=0,1)$ are the orthonormal basis in Cartan subalgebra of
$\dof$ and $k_i=q^{\frac{\eps_i - \eps_{i+1}}{2}}, \eps_{\iota+2} = \eps_i.$

The comultiplication (\ref{eq:4}) converts into (\ref{eq:3}) by changing:
$z \cdot e_i \Rightarrow e_i, \\ z^{-1} \cdot f_i \Rightarrow f_i, \ z^2 z_i
\Rightarrow z_i,$
where $z:=q^{\frac{\eps_0 + \eps_1}{2}}$ lies in the center of $\dof$.

{}From the representation of finite dimensional quantum algebra the
pa\-ra\-metri\-zed
representation of its affine extension can be constructed. Let's regognize this
method for $\dof$. If $e, f, k$ are the generators of $\dof$, then there is the
homomorphism $\rho_x : \dof \Rightarrow \dom$ \cite{2}:
\be
\begin{array}{lll}
\rho_x (e_0) = x \cdot f & \rho_x (f_0) = x^{-1} \cdot e & \rho_x
(k^{\pm 1}_0) = k^{\mp 1} \nonumber \\
\rho_x (e_1) = e & \rho_x (f_1) = f & \rho_x (k^{\pm 1}_1) = k^{\pm 1}
\end{array}
\label{eq:6}
\ee
This homomorphism converts representations of $\dom$ to parametrized
representation of the affine quantum algebra $\dof$ with the central charge
 equal to zero.

The Hopf algebra structure allows to consider the action of $\dof$ on thensor
products of representations. Let $\pi_1$ and $\pi_2$ be representations of
$\dom$ on $V_1$ and $V_2$ respectively. Then, as it was mentioned above, $\pi_i
(x_i):=\pi_i \circ \rho_{x_i}, i=1,2$ are the representations of $\dof$. The
 equivalence of tensor products $\pi_1 (x_1) \otimes \pi_2 (x_2)$ and
$\pi_2 (x_2) \otimes \pi_1 (x_1)$ implies a relation between parameters of the
representations and means the existence of some intertwining operator
$\hat{R}_{\pi_1 \pi_2} (x_1, x_2)$ from $V_1 \otimes V_2$ into $V_2 \otimes
V_1$ such that \cite{1}):
\bea
&& \hat{R}_{\pi_1 \pi_2} (x_1,x_2) \pi_1 (x_1) \otimes \pi_2 (x_2) (\Delta(g))
\nonumber\\
&& = \pi_2 (x_2) \otimes \pi_1 (x_1) (\Delta(g)) \hat{R}_{\pi_1 \pi_2} (x_1,
x_2),
\label{eq:7}
\ena
where $ g \in \dof$. We can take $x_2 =1, x_1 = x$ because $\hat{R}$ depends on
$\frac{x_2}{x_1}$ only. Then the equations (\ref{eq:7}) for $g= f_0, f_1$ can
be represented in the following form:
\bea
\hat{R} (x) \cdot (f \otimes k^{-1} + z^{-1}_1 \cdot k \otimes f) &=& (f
\otimes
k^{-1} + z^{-1}_1 \cdot k \otimes f) \cdot \hat{R} (x) \nonumber \\
\hat{R} (x) \cdot (xf \otimes k + z_0 \cdot k^{-1} \otimes f) &=& (f \otimes
k + xz_0 \cdot k^{-1} \otimes f) \cdot \hat{R} (x)
\label{eq:8}
\ena
{}From (\ref{eq:8}) we obtain:
\bea
\hr(x) \cdot (1 \otimes f) &=& [\Delta(f_1) \cdot \hr(x) \cdot x (1 \otimes
k^2)- \Delta (e_0) \cdot \hr(x) ] \nonumber \\
&& \times (xz^{-1}_1 \cdot k \otimes k^2 - z_0
\cdot k^{-1} \otimes 1)^{-1} \nonumber \\
\ & \ \nonumber \\
\hr(x) \cdot (f \otimes 1) &=& [\Delta(f_1) \cdot \hr(x) \cdot (z_0 z_1
\cdot  k^{-2}
\otimes 1)- \Delta (e_0) \cdot \hr(x) ] \nonumber \\
&& \times (z_0 z_1 \cdot k^{-2} \otimes k^{-1} - x( 1 \otimes k))^{-1},
\label{eq:9}
\ena
where for simplicity we use the notations:
\[
\Delta (f_i):=\pi_2 (1) \otimes\pi_1 (x) \left( \Delta(f_i) \right), \Delta
(e_i):= \pi_2
(1) \otimes \pi_1 (x) \left( \Delta (e_i) \right).
\]

If $\pi_1$ and $\pi_2$ are the highest weight representations with highest
vectors $v^{(1)}_c$ and $v^{(2)}_c$ respectively and $V_1 \otimes V_2$
decomposes into direct sum of pairwise nonequivalent irreducible
representations, then $\hr(x) (v^{(1)}_c \otimes v^{(2)}_c) = \alpha \cdot
(v^{(2)}_c \otimes v^{(1)}_c)$, where $\alpha \in {\bf C}$ can be equated to
1. In this case we can use
(\ref{eq:9}) to obtain the recursive formula for $\hr(x)(f^{r_1} v^{(1)}_c
\otimes f^{r_2} v^{(2)}_c)$. In fact, using the equalities:
\bea
\left( 1\otimes k^2 \right) \left( f^{r_1} v^{(1)}_c \otimes f^{r_2} v^{(2)}_c
\right) &=& \lambda_{(2)} \
q^{-2r_2} \left( f^{r_1} v^{(1)}_c \otimes f^{r_2} v^{(2)}_c \right) ,
\nonumber \\
\left( k^{-2} \otimes 1 \right) \left( f^{r_1} v^{(1)}_c \otimes f^{r_2}
v^{(2)}_c \right) &=&
\lambda^{-1}_{(1)} \ q^{2r_2} \left( f^{r_1} v^{(1)}_c \otimes f^{r_2}
v^{(2)}_c \right) ,
\nonumber
\ena
we obtain from (\ref{eq:9}):
\bea
&& \hr(x) (f^{r_1} \otimes f^{r_2}) \vv = \left[ x\lambda^2_{(2)} q^{-2(r_2-1)}
\Delta(f_1) - \Delta (e_0) \right] \hr(x) \nonumber \\
&& \times (f^{r_1} \otimes f^{r_2-1})(q^{-2r_2 -r_1 -2} xz_1^{-1} k \otimes k^2
-q^{r_1} z_0 k^{-1} \otimes 1)^{-1} \vv \nonumber \\
\ && \ \\
&& \hr(x) (f^{r_1} \otimes f^{r_2}) \vv = \left[ \lambda^{-2}_{(1)} \
q^{2(r_1-1)}
(z_0 z_1 \otimes 1)\Delta(f_1) - \Delta (e_0) \right] \nonumber \\
&& \times \hr(x)(f^{r_1-1} \otimes f^{r_2})(q^{2r_1 +r_2 -2} z_0 z_1 k^{-2}
\otimes k^{-1} -q^{-r_2} x \cdot 1 \otimes k)^{-1} \vv
\nonumber \label{eq:9'}
\ena
Here $\lambda_{(i)} (i=1,2)$ are values of $k$ on highest vectors  $v^{(i)}_c$.
Using this and denoting by $z^{(i)}_j$ the values of central elements $z_j
(j=0,1)$ on $V_i$, we obtain by induction from (\ref{eq:9'}):
\bea
R(x) \left( f^{r_1} \right. & \otimes & \left. f^{r_2} \right) \vv
= \frac{(z_1^{(1)})^{-r_1}}
{\prod_{\iota=0}^{r_1 + r_2-1} \left( q^\iota (\lambda_1 \lambda_2)^{-1}
z^{(1)}_0 - q^{- \iota}
(\lambda_1 \lambda_2) xz^{(1)-1}_c \right) } \nonumber \\
&\times& \prod^{r_1 -1}_{\iota_1 =0} \left[ \lambda^{-1}_{(1)} q^{\iota_1}
z^{(2)}_c z^{(2)}_1
\Delta (f_1) - \lambda_{(1)}  q^{- \iota_1} \Delta(e_0) \right] \nonumber \\
&\times& \prod^{r_2 -1}_{\iota_2 =0} \left[ \lambda^{-1}_{(2)}
q^{\iota_2} \Delta(e_0) - x\lambda_{(2)} q^{- \iota_2} \Delta(f_1) \right]
(\vvv)
\label{eq:10}
\ena

It can be proved that this $R$-matrix commutes also with $\Delta(f_0)$ and
$\Delta(e_1)$ (As in example 3 the condition on parameters of representations
follows from the equality of central elements in tensor products).

The formula (\ref{eq:10}) generalizes the results of \cite{4} in
case of spectral parameter depending $R$-matrix and any highest weight
representation of quantum algebra $\dom$. In the particular case of
$x=1, z_i=1$,
the formula (\ref{eq:10}) determines the reflection symmetric intertwiner for
$\dom$, constructed in \cite{4}. Note, however, that we used generators,
slightly differing by ones of $\dom$, so there are some differences in
formulas. We kept track by the method of \cite{4} in our derivation.

Consider now some examples.

\indent

{\bf Example 1.} First we consider the case of general $q$. We put $z_i=1$. The
irreducible representations of quantum algebra for general $q$ are the
deformations of representations of corresponding classic algebra and
characterized by half integer highest weight. As it was proved by in ref.
\cite{3}, $\hr(x)$ exists and is unique. It also satisfies Yang-Baxter
equations on
$\pi_1 (xy) \otimes \pi_2 (x) \otimes \pi_3 (1)$ \cite{1,2}:
\be
(\hr(x) \otimes id)(id \otimes \hr(xy) (\hr(y) \otimes id) = (id \otimes
\hr(y)) (\hr(xy) \otimes id) (id \otimes \hr(x))
\label{eq:11}
\ee

{}From the uniqueness of $R$-matrix it follows that constructed above operator
(\ref{eq:10}) coincides with Jimbo's $R$-matrix in \cite{2}. The later, which
had been represented by means of projecting operators, can be also represented
in the explicit form:
\bea
\hr(x) && v^{(j_1)}_{r_1} \otimes v^{(j_2)}_{r_2} = \prod^{r_1+ r_2 -1}_{\iota
=0} \left( q^{\iota -j_1 - j_2} -xq^{-(\iota -j_1 -j_2)} \right) ^{-1}
\nonumber \\
&&  \times \prod^{r_1 -1}_{\iota_1 =0} \left[ q^{\iota_1 - j_1}
\Delta(f_1)-q^{-(\iota_1 -j_1)} \Delta(e_0) \right] \nonumber \\
&& \times \prod^{r_2 -1}_{\iota_2 =0} \left[ q^{\iota_2 - j_2}
\Delta(e_0)-xq^{-(\iota_2 -j_2)} \Delta(f_1) \right] \left( v^{(j_2)}_c \otimes
v^{(j_1)}_c \right),
\label{eq:12}
\ena
where $v^{(j)}_k := f^k v^{(j)}_c$.

\indent

{\bf Example 2.} Let $q$ be a root of 1 and $N$ be the minimal integer, such
that $q^N =1$. For simplicity we consider the case of odd $N$ only. Then $e^N,
f^N, k^N$ lie in the center of $\dom$ \cite{10,11}. It
follows from this that
in this case appear new irreducible representations that are called periodic or
cyclic \cite{10,11}. They are $N$-dimensional and parametrized by
3 complex numbers. The factorized $S$-matrix of $sl_q(N, {\bf C})$-generalized
chiral Potts model has been found first by Bazhanov et all in ref.\cite{16}.
In the article \cite{6} Date, Jimbo, Miki and Miwa by means of \cite{16}
constructed intertwining operators
(\ref{eq:7}) for the minimal cyclic representation of the
$\hat{g}l_q (N, {\bf C})$ affine algebra with central extension. They also
proved the Yang-Baxter equations between them for parameters of corresponding
representations lying on some
algebraic curve.

Recall that the minimal cyclic representation of $\dom$ has the following form:
\be
\begin{array}{ll}
\pi_\xi(e)w_m = yq^{\mu_1 + \mu_0} [2\mu_1 -m]_q \ w_{m-1} & \pi_\xi (k) w_m =
q^{\mu_1 - \mu_0 -m-1} w_m \nonumber \\
\ & \ \\
\pi_\xi(f)w_m = y^{-1}q^{-\mu_1 - \mu_0} [2\mu_0 +m+2]_q \ w_{m+1} & \pi_\xi
(z_i) = \frac{c_i}{c_{i+1}} (i=0,1), \nonumber
\label{eq:13}
\end{array}
\ee
where $c_2:=c_0, w_N:=w_0, m=0,1...N-1$, and $\xi:= \left( q^{\mu_0},
q^{\mu_1}, y, \frac{c_1}{c_2} \right)$ is a parameter of the representation.

After affinization of (\ref{eq:13}) and imposing comultiplication (\ref{eq:3})
we obtain minimal representation of $\dof$, considered in \cite{6}. (Here we
use
slightly different basis and comultiplication). Recall that inertwining
operator: $V_\xi \otimes V_{\xi}'  \Rightarrow V_{\xi}' \otimes
V_\xi$ exist if $\xi$ and ${\xi}'$ lie on algebraic curve, which is
parametrized by 2 complex varieties $s$ and $s'$:
\be
\begin{array}{lll}
u^N_i = s - \lambda_i & v^N_i = s - \mu_i & \\
\ & \ \\
u'^N_i = s'_i - \lambda_i & v'^N_i = s' - \mu_i & (i=0,1),
\nonumber
\end{array}
\label{eq:14}
\ee
where:
\[
\frac{u_i}{u'_i} = \frac{q^{2\mu_i}}{c_i},  \ \frac{v'_{i-1}}{v_{i-1}} =
q^{2\mu_i} c_i, \ x_i = \omega_i^{\frac{1}{N}} \frac{u'_i}{v'_i},
\]
(Here $x_1:=y, x_0:= \frac{x}{y}$). The moduli parameters $\gamma:=(\omega_i,
\lambda_i,\mu_i)$ are fixed for different representations ($\mu_i$ is not
the same variable as in (\ref{eq:13})).

If $\mu_1 =0$ then $\pi_\xi (e) w_0=0$, and periodic representation converts
into semiperiodic one. It follows from (\ref{eq:14}) that $\mu_0 = \lambda_1$
in
$\gamma$ in this case. So, the $R$-matrix, which intertwines semiperiodic
representations, is a particular case of $R$-matrix, intertwining periodic
ones. \footnote{We use the unicity arguments of $R$-matrix which follows from
the irreducibility of tensor product in $\dof$} (For $\mu_0 = \lambda_1$, as it
can be verified, there are no
singularities in $R$-matrix.) Thus the formula (\ref{eq:10}) for $R$ matrix can
be used for semiperiodic representations. In the case of $x=c_i=z_i=1$ we
obtain
$R$-matrix, considered in \cite{4}.

\indent

{\bf Example 3.} Consider now the tensor product $V_\xi \otimes V_j$, where by
$V_\xi$ we denote the semiperiodic representation with parameter $\xi=(q^\mu,
\lambda$) and by $V_j$ the spin $j$ one, $2j+1<N$. We take the following basis
for $V_\xi$ \cite{12}:
\be
\begin{array}{ll}
\pi_\xi(e) w_m^{(\xi}) = [2\mu-m+1]_q [m]_q w_{m-1} & \pi_\xi(k) w^{(\xi)}_m =
q^{\mu-m} w_m \\
\ & \ \\
\pi_\xi(f) w^{(\xi)}_m = w_{m+1}, m=0,..N-2 & \pi_\xi(f) w_{N-1}^{(\xi)} =
\lambda w_0
\label{eq:14'}
\end{array}
\ee

We shall show that $\hr(x)$, obtained using (\ref{eq:10}) for all
$z^{(i)}_j=1$, is an intertwiner from $V_\xi \otimes V_j$ into $V_j \otimes
V_\xi$ of $\dof$ (see (\ref{eq:7})).

First we note that $\hr(x)$ commute with $\Delta(f_1)$ and $\Delta(e_0)$. To
see this it's enough to check:
\be
\frac{\prod^{N-1}_{\iota=0} \left[ q^{\iota - \mu} \Delta(f_1)-q^{-(\iota -
\mu} \Delta (e_0) \right] (v^{(j)}_c \otimes v^{(\mu)}_c)}{\prod^{N-1}
_{\iota=0} \left( q^{\iota-\mu-j} -xq^{-(\iota -\mu-j)} \right)}
= \lambda (v_c^{(j)} \otimes v^{(\mu)})_c
\label{eq:15}
\ee
\be
\frac{\prod^{2j}_{\iota=0} \left[ q^{\iota - j} \Delta(e_0)-xq^{-(\iota -
j)} \Delta (f_1) \right] (v^{(j)}_c \otimes v^{(\mu)}_c)}{\prod^{2j}
_{\iota=0} \left( q^{\iota-\mu-j} -xq^{-(\iota -\mu-j)} \right)} = 0
\label{eq:16}
\ee
(\ref{eq:15}) can be proven using commutativity of $\Delta(f_1)$ and
$\Delta(e_0)$, Gauss binomial formula:
\[
\prod^{N-1}_{\iota=0} (1-q^{2\iota} a) = \sum^N_{\iota=0} (-1)^\iota \left[
\begin{array}{c} N \\ 1 \end{array} \right] _q q^{\iota (\iota - 1)} a^\iota \
(=1+a^N \mbox{for our} \ q),
\]
the equations $\Delta(f_1)^N = f^N \otimes k^{-N} + k^N \otimes F^N,
\Delta(e_0)^N = f^N \otimes k^N + x^N k^{-N} \otimes f^N$, the fact that $f^N
|_{V_j} \equiv 0, \ k^{2N} |_{V_j} \equiv id$ and (\ref{eq:14}). To prove
(\ref{eq:16}) we consider it for general $q$. Because of existence and
uniqueness of an $R$-matrix for half integer values of $\mu$ the equation
(\ref{eq:16}) is valid for such $\mu$. It is evident, that the numerator in
(\ref{eq:16}) can be represented as a linear combination of $v^{(j)}_p \otimes
v^{(\mu)}_m, p+m=2j+1$, with the coefficients, which are $\mu$-independent
polynomials on $q^\mu$. Because of vanishing such polynomials on infinite
numbers of points $q^{half integer}$ they are vanished trivially. So,
(\ref{eq:16}) is valid.

Now recall the fact that $V_\xi \otimes V_j$ as a $\dom$ module is a fully
reducible. It decomposes into $2j+1$ semiperiodic representations $V_\xi(i),
i=1..2j+1$ (\cite{12}). From this it follows that $\Delta(e_1)$ and
$\Delta(f_0)$ also commute with $\hr(x)$. Indeed, if $v^{\mu_i)}_c$ is highest
weight of $V_\xi(i)$ then considering the half integer values of $\mu$ as above
it can be proven that $\Delta(f_0) \hr(x)  v^{(\mu_i)}_c = \Delta(e_1) \hr(x)
v^{(\mu_i)}_c =0$. Then:
\[
\Delta(e_1) \hr(x) \Delta (f^k_1) v^{(\mu_i)}_c  = \Delta(e_1) \Delta(f^k_1)
\hr(x) v^{(\mu_i)}_c  = \hr(x) \Delta (e_1) \Delta(f_1) v^{(\mu_i)}_c
\]
\[
\Delta(f_0) \hr(x) \Delta (e^k_0) v^{(\mu_i)}_c  = \Delta(f_0) \Delta(e^k_0)
\hr(x) v^{(\mu_i)}_c  = \hr(x) \Delta (f_0) \Delta(e_0) v^{(\mu_i)}_c
\]
So, we proved the $\hr(x)$ is an intertwiner from $V_\xi \otimes V_j$ to $V_j
\otimes V_\xi$ of affine $\dof$. The Yang-Baxter equations on $V_\xi \otimes
V_j \otimes V_j$ can be proven using the  irreducibility of tensor product
as in \cite{6}.

At the end we write the components of $\hr(x):V_\xi \otimes V_j \Rightarrow
V_j \otimes V_\xi$ for $N=3, j= \shalf$.
\bea
\hr^{00}_{00} (x) &=& 1 \ \  \ \ \ \ \ \hr^{10}_{10}(x) = -
\frac{[2\mu]}{(-\mu - \shalf)_x}
\ \ \ \ \ \ \ \hr^{01}_{10}(x) = \frac{(-\mu + \shalf)_x}{(- \mu - \shalf)_x}
\nonumber
\\
\hr^{10}_{01}(x) &=& \frac{(\mu - \shalf)_x}{(- \mu - \shalf)_x} \ \ \ \ \ \ \
\hr^{01}_{01}
(x) = \hr^{02}_{11}(x)= - \frac{x}{(- \mu - \shalf)_x} \nonumber \\
\hr^{11}_{11}(x) &=& \frac{(\mu - \sm{3}{2})_x}{(- \mu - \shalf)_x}
\ \ \ \ \ \ \ \hr^{02}_{20}(x) = \frac{x(1)_x}{(- \mu - \shalf)_x
(-\mu + \sm{3}{2})_x}
\nonumber \\
\hr^{11}_{20}(x) &=& \frac{x[\mu - \sm{3}{2}] + (\mu - \shalf)_x (1-x)}
{(- \mu - \shalf)_x  (-\mu + \shalf)_x} \ \ \ \ \ \ \
\hr^{00}_{21}(x) = \frac{\alpha x}
{(- \mu - \shalf)_x} \nonumber \\
\hr^{12}_{21}(x) &=& \frac{x[2\mu - 3] + (\mu - \sm{3}{2})_x (-\mu+ \shalf)_x}
{(- \mu - \shalf) (- \mu + \sm{3}{2})_x}
\nonumber
\ena
Here we omitted the index $q$ in $[n]_q$ and used the notation:
\[
(n)_x := \frac{q^n - xq^{-n}}{q-q^{-1}}
\]

We would like to thank V.Bazhanov for a comments.

\indent

{\bf Note added}. When this work was finished and submited for a publication
B.M.McCoy kindly informed us that similar results was obtained by I.T.Ivanov
and D.B.Uglov in the article Phys.Lett. A 167 (1992) 459, for which we express
 our gratitudes to him.

\end{document}